\documentclass[12pt]{article}

\usepackage{scicite}

\usepackage[ final ]{graphicx}

\usepackage{times}

\newenvironment{sciabstract}{%
\begin{quote} \bf}
{\end{quote}}

\newcounter{lastnote}

\title{A self-referenced single-electron quantized-current source}

\author{Lukas Fricke,$^{1\ast}$ Michael Wulf,$^{1\dagger}$ Bernd Kaestner,$^{1}$ Frank Hohls,$^{1,\ddagger}$\\
Philipp Mirovsky,$^{1}$ Brigitte Mackrodt,$^{1}$ Ralf Dolata,$^{1}$ Thomas Weimann,$^{1}$\\
Klaus Pierz,$^{1}$ Uwe Siegner,$^{1}$ Hans W. Schumacher,$^{1}$\\
\\
\normalsize{$^{1}$Physikalisch-Technische Bundesanstalt,}\\
\normalsize{Bundesallee 100, D-38116 Braunschweig, Germany}\\
\normalsize{$^\ast$To whom correspondence should be addressed; E-mail: lukas.fricke@ptb.de.}\\
\normalsize{$^\dagger$Author is deceased}\\
\normalsize{$^\ddagger$To whom correspondence should be addressed; E-mail: frank.hohls@ptb.de.}
}

\date{}

\begin{document} 


\baselineskip24pt


\maketitle


\begin{sciabstract}
The future redefinition of the international system of units in terms of natural constants requires a robust, high-precision quantum standard for the electrical base unit ampere. However, the reliability of any single-electron current sources generating a nominally quantized output current $I = ef$ by delivering single electrons with charge $e$ at a frequency $f$ is eventually limited by the stochastic nature of the underlying quantum mechanical tunnelling process. We experimentally explore a path to overcome this fundamental limitation by serially connecting clocked single-electron emitters with multiple in-situ single-electron detectors. Correlation analysis of the detector signatures during current generation reveals erroneous pumping events and enables us to determine the deviation of the output current from the nominal quantized value $ef$. This demonstrates the concept of a self-referenced single-electron source for electrical quantum metrology. 
\end{sciabstract}

The field of quantum metrology promises measurement techniques and standards providing highest precision and universality by making reference to fundamental constants of nature~\cite{Planck1900}. For electrical  quantum metrology two macroscopic quantum effects - the Josephson effect~\cite{Josephson1962} and the Quantum-Hall effect~\cite{Klitzing1980} - have become the basis of robust and versatile quantum standards for the units Volt and Ohm, respectively. The anticipated redefinition of the international system of units (SI) based on quantum metrology, however, requires a quantum standard for the SI base unit Ampere~\cite{CCEM2009,BIPM2011}. As candidates for such a quantum current standard single-electron pumps and turnstiles \cite{Geerligs1990,Pothier92,Keller1996,Keller1999,Pekola2008,Faist2011,Roche2013,Blumenthal2007,Kaestner2008PRB,Faist2011,Giblin2012} (see as a review,e.g. \cite{Pekola2013}) have been explored. Among them dynamic semiconductor quantum dots (QDs) \cite{Blumenthal2007,Kaestner2008PRB,Giblin2012} have recently demonstrated very promising characteristics. However, the operation of all of these kinds of devices is not based on a robust macroscopic quantum effect, but on a quantum mechanical tunnelling process during the periodic capture of a charge $e$ from a source reservoir \cite{Zimmerman2004,Kashcheyevs2010,Fricke2013}. The stochastic nature of this process~\cite{Maire2008} inevitably leads to small random deviations from the ideally quantized current of $I = ef$, with $f$ being the pumping rate, hindering the realization of a robust quantum current source with lowest uncertainty.

As an alternative approach towards quantum current metrology the detection and counting of electrons passing randomly through a nanostructure has been explored and demonstrated both for unipolar~\cite{Bylander2005} and bipolar~\cite{Fujisawa2006} currents. However, the sensitivity and bandwidth of present single-charge detectors places severe limits on the current amplitude and achievable uncertainty of this approach. 
The late M. Wulf has recently proposed~\cite{Wulf2013} a way to overcome these two fundamental problems of quantum-current metrology by combining the two concepts of single-electron pumping and single-electron detection in a single device: the so-called self-referenced single-electron current source. In this quantum circuit the nominally quantized single-electron current is generated in a serial arrangement of single-electron pumps and single-electron detectors~\cite{Fricke2011}. During self-referenced operation only the much less frequently occurring stochastic pumping errors are detected instead of detecting all charges passing the circuit. Doing so the deviations from the nominal current $I = ef$ are reliably determined and accounted for allowing the realization of a low-uncertainty single-electron quantum current source with validated output current. However, such self-referenced current generation has not been experimentally demonstrated yet. 

Here, we present a quantum circuit integrating QD-based single-electron pumps and single-charge detectors to implement all basic features of a self-referenced single-electron current source. We demonstrate single-electron pumping and in-situ charge detection during single-electron current generation at low frequencies. It is shown that single-electron error detection combined with statistical analysis yields corrected values of the quantized current with uncertainties that are reduced by more than one order of magnitude as compared to an individual single-electron pump. Quantum circuits of the type presented in this paper should allow scaling to higher currents and lower uncertainties for future applications as a robust quantized-current standard for the SI base unit ampere.

The device under investigation is a hybrid semiconducting-metallic nanostructure as shown in Fig.~1~\cite{Fricke2013}. From the lower left to the upper right a semiconducting channel is formed by etching a GaAs/AlGaAs heterostructure with a high-mobility two-dimensional electron gas (2DEG) situated $90$~nm underneath the surface. The resulting 2DEG channel consists of source and drain (green) contacts and intermediate nodes 1,2 (blue, red) each separated by one of the three single-electron pumps P1-P3. Each pump is defined by three metallic top gates (yellow) crossing the channel. The electrical potentials of the nodes and hence the nodes' charge states are monitored by two single-electron transistors (SET)~\cite{Fulton1987,Kuzmin1989} operated as electrometer detectors labelled D1 (blue) and D2 (red), respectively. The SETs are fabricated by two-angle shadow evaporation of aluminium~\cite{Dolan1977}. For enhanced capacitive coupling of the detectors to the nodes gates with floating potentials (FG, yellow) are deposited. The SET detectors are operated at fixed bias voltage and the current, acting as detector signal, is measured by synchronized digitizer cards at a sampling rate of $12$~kS/s after digital averaging. Doing so we oversample the detector response by about 20 times. The measurements presented are all performed in a dry dilution cryostat at nominal base temperature of about $25$~mK. 

The working principle of the pump has been discussed in detail in Refs.~\cite{Kaestner2008APL,Kaestner2008PRB,Kashcheyevs2010,Leicht2010,Fricke2013}. The robust operation of these electron pumps in a serial arrangement has been demonstrated in Ref.~\cite{Fricke2011} and has been  adapted for this work as follows: By applying negative voltages of about $-200$~mV to the first two gates of each pump (entrance and exit gate, respectively) isolated quantum dots (QD) are formed between the gates. (The third gate of each pump is grounded and not used in our experiments.) 
For these applied voltages the QD ground state is well above the chemical potential of the semiconductor channel and the QD is empty. To induce pumping of a single electron, a single, cosine-shaped pulse with frequency $f=40$~MHz and amplitude $V=-75$~mV at the output of the generator is superimposed onto the entrance gate of pump $i$. During the pulse the entrance barrier is first lowered to allow tunnelling onto the QD from the source and then raised back up. In this cycle $n_i$ electrons are captured from the pump's source side in the QD with probability $P^{(i)}_n$ and subsequently emitted to the drain side. Note that whenever no pump pulse is applied the high gate barriers of the QD prevent electrons from tunnelling between the nodes.  

A sketch of the pulse shape as well as the pulse sequence applied to the pumps during this work is shown in Fig.~2A. 
The pulses are delayed by $\tau=20$~ms each. The delay $\tau$ is about ten times larger than the response time of our detectors $\tau_d\approx 1.5$~ms. Hence the resulting charge change following each pumping event can be reliably resolved by the detectors.
The pulses of the sequence can be separated in two groups. The first three pulses ((i)-(iii)) constitute what we refer to as a marker sequence. Using this pulse pattern, we pump $n = 1 $ additional electron sequentially through the structure from source to node 1, to node 2, and to drain by subsequent application of a pump pulse to P1, P2, and P3. This allows for a calibration of D1 and D2 in terms of the signature of $n = 1 $ (or more) additional electrons on node 1 and 2 during the measurement (see supplemental material). Additionally, we can extract the individual pump characteristics from these pulses for every pump $i$ i.e. the probability $P^{(i)}_n$ of pumping $n_i$ electrons per pumping cycle (see supplemental). The last two elements (pulses (iv)-(v)) of the pulse sequence contain two sets of pulses where all three pumps are triggered simultaneously. Here, we change the detection scheme from an absolute measure of number of transferred electrons to the detection of pumping errors as described below.  

The expected detector signals D1 (blue) and D2 (red) during one sequence are sketched at the bottom of Fig.~2A. Here we assume that all pumps transfer exactly one electron per pulse. Horizontal dashed lines indicate the two detector states $d_{(1,2)},d_{(1,2)}+1$ of node $(1,2)$ reflecting a change in the number of electrons on the corresponding node by one electron. When triggering pump P1 (pulse (i)), one electron is transferred from the source lead to node 1, leading to a step in D1 while D2 remains unaffected. By the following pulse (ii) one electron is transferred across P2 from node 1 to node 2, resulting in a step in both detectors with opposite sign. D1 returns to the initial value $d_1$ and D2 increases to $d_2 + 1$. Lastly in the marker sequence, pump P3 is triggered (iii) removing one electron from node 2 to drain with D2 returning to its initial state. In the following synchronized pumping of P1-P3 (pulses (iv)-(v)), D1 and D2 both remain in their initial state as one electron is pumped on to and off each node leaving the node charge unchanged. The same behaviour is also observed in the experimental data of error free pumping operation shown in Fig.~2B. Here the measured signals of D1 (blue) and D2 (red) are shown during five consecutive pumping sequences (1-5) as described above. The coloured vertical lines indicate the different pulses (i) - (iv) of the sequence as marked in Fig.~2A (see dashed lines connecting Figs.~2A,B). The high signal-to-noise ratio of the detector signals allows to distinguish the two charge states $d_{(1,2)},d_{(1,2)}+1$ of both nodes reliably and thus to follow the charge transfer during the marker sequences.

As discussed above, stochastic tunnelling errors can occur during single-charge pumping. If so, it is possible to identify and attribute pump errors with our device during sequential as well as during synchronous operation as shown exemplary in Figs.~2C-F (black arrows). During the marker sequence the transferred charges are measured directly. Fig.~2C shows the measured detector signals of an event where P1 fails to pump an electron during pulse (i) (black arrow). As a consequence D1 and D2 remain in the same state before and after pulse (i). However, during pulse (ii) P2 transfers an electron from node 1 to node 2 thereby shifting the baseline of detector D1 by $-1e$. In the same manner pumping errors of P2, P3 in the marker sequence can be detected reliably and analyzed statistically due to their characteristic signature (not shown).

Also during synchronous operation (pulses (iv),(v)) pumping errors can be detected and analyzed. Detector traces representing examples of such errors are shown in Figs.~2D-F. In Fig.~2D the D1 signal shows a sudden drop indicating that the charge of node 1 is reduced by $1e$. In contrast the charge of node 2 remains constant as expected. This signature is most likely the consequence of P1 missing to pump an electron while P2 and P3 are operating properly during synchronous pumping. If pump P2, connecting both nodes, misses a pump cycle, this should result in a simultaneous change in the signals of D1, D2 with opposite sign: The charge of node 1 increases by $1e$ whereas the charge of node 2 is lowered by $1e$. Such a signature is indicated by the arrow in Fig.~2E. In case of pump P3 missing a cycle, node 1 should remain unaffected and only the detector D2 should detect an additional electron on node 2. Such a trace is shown in Fig.~2F. 

This data clearly shows that it is possible to detect pumping errors during synchronous operation of the three single electron pumps connected in series. However, note that the attribution of error signatures to failures of the individual pumps P1-P3 is not unique. As an example, the signature of perfect series operation with both detectors at constant level could also result from simultaneous failure of all pumps~\cite{Wulf2013}. To illustrate this intrinsic problem of the data analysis in more detail, we analyze the different scenarios leading to the signature of Fig.~2F in more detail. Fig.~3A again shows the detector signal of the event (left) and a schematic of the node charge (right): D1 shows a constant number of electrons ($d_1$) on node 1 while D2 indicates an additional electron ($d_2+1$) on node 2. Below the schematic a table contains three possible scenarios of charge transfer and their corresponding probabilities which are compatible with the detector signature. This probability vector for transferring $n$ electrons across P3 is derived from the characterization of the pumping statistics of each pump by analyzing the marker sequence (see supplemental material for details). As mentioned above, the most likely explanation of this signal is a missing pump event by pump P3 (first line). This scenario is the most probable one with probability of 0.9999. But also a simultaneous error by the other two pumps P1 and P2, both transferring two electrons, results in this signature. Considering the working points of the pumps, this coincidental error of P1 and P2 is quite unlikely with a probability of $10^{-4}$ (second line). The next-order process leading to the same charge signals involves erroneous pumping by all three pumps transferring $n_{(1,2,3)}=(3,3,2)$ (line 3) and is already very unlikely with a probability of the order of $10^{-9}$. Generally speaking, the probability of an event scales inversely with the number of failing pumps, i.e., the fewer pumps required to deviate from their nominal working point in order to explain the signature, the higher the probability. The same argument holds for the introductory example: For the given device and operation parameters the probability of all pumps failing in synchronous operation instead of transferring one electron is $1.5\cdot 10^{-5}$. Such small probabilities of higher-order processes will generally lead to a minor broadening of the final probability distribution and hence to a small increase of the uncertainty of the current generation which is determined from the width of the probability distribution while the maximum of the distribution yields the corrected value of the current.

We also include in our analysis that for the given device and operation parameters specific events can occur which cannot be identified with high reliability. Fig.~3B shows an event leading to the signature $(d_1+1,d_2)$. Here, two possible explanations for the observed signature are of almost same probability as given in the table. These are on the one hand a coincidental error by the last two pumps missing the transfer cycle, on the other hand an error by the first pump transferring an additional electron. Due to the higher probability of the latter event, we will assume one electron being transferred across P3. Due to the large probability of a misattribution, the corresponding lack of knowledge leads directly to an abrupt broadening of the electron number distribution and hence of the uncertainty of the output current. In an optimized device events of this kind should be avoided by optimizing the working points of the individual pumps or by increasing the number of pumps connected in series.

By applying statistical analysis to a long series of pumping events the output current and its full uncertainty distribution can be obtained. The current output of the device is derived from the number of electrons transferred across P3 to drain. Two examples of such analysis are shown in Fig.~3C. Here, the deviations of the analyzed current output from the nominal quantized current with $n_3=1$ (i.e. the difference of transferred electrons to the number of pulses applied to pump P3) is shown as a function of the pulse index. The traces are offset by $-10$ for clarity, their length is limited by the requirement of both detectors being sensitive at the same time. The limitation in dynamic range of the detectors can be lifted in future by e.g. applying feedback to the SETs' gates to stabilize them at optimal working points. Each jump in the traces in Fig.~3C reflects a deviation from one electron per pulse across P3: Downwards corresponds to a missing electron, upwards to a surplus electron at this specific pulse. In the upper, green trace in total, nine electrons less than the nominal number are transferred across P3 ($2.1\%$). In the lower blue trace, a total of seven electrons ($1.3\%$) are missing.

Yet, as discussed before, not only the absolute number of transferred electrons is obtained, but also the full probability distribution of the different possible scenarios and hence the full uncertainty of the generated single-electron current output is derived. This statistical analysis of the output current for both traces of Fig.~3C are shown in Figs.~3D-E. The probability distribution in Fig.~3D after accounting of the green trace in Fig.~3C has significantly narrowed compared to the projected probability distribution of pump P3 (grey, dashed lines) without self-referencing. However, during the measured trace several detector events with scenarios of comparable probability (e.g. of the type discussed in Fig.~3B) result in a distribution of still considerable current uncertainty. The expectation value resulting from the distribution is $405.2\pm 1$, i.e. we expect after error accounting to transfer 405 electrons. Without accounting (grey distribution), we expect to transfer most likely $399\pm 4.1$ electrons with the uncertainty given in both cases by the square-root of the variance of the corresponding probability distribution. Hence, error accounting has improved the uncertainty of the output current, but the sub-optimal working point limits the accuracy gain.

In contrast, in the trace shown in Fig.~3E the working point was chosen such that only distinguishable errors occurred. This results in a very narrow probability distribution after error accounting (blue bar diagram, solid edge) compared to the expected probability distribution for the individual pump P3 based on the experimentally determined pump probabilities (grey bars, dashed edges). After error accounting, we know that $524$ electrons have been transferred with a probability of $99.4\%$ by $531$ pulses onto P3 (electron number uncertainty $\pm 0.08$), whereas for the individual pump P3 we expect on average $519$ electrons with an uncertainty of $\pm 3.9$. The output current in this trace equals to an average current of $I=4.743$~aA over a time interval of $17.7$~s with a resulting uncertainty of $0.7\cdot 10^{-21}$~A after accounting. Compared to the individual pump P3, the uncertainty is reduced by a factor of about $50$ and thus by more than one order of magnitude. 

In the future, the reduction of the output uncertainty of a such self referenced single electron pump could be significantly enhanced in two different ways: On the one hand, the pumps themselves can be tuned to highly asymmetric transfer rates making, for example, an event including the transfer of two electrons by any of the pumps very unlikely. On the other hand, the addition of further pumps and detectors in the serial arrangement will exponentially reduce the probability of correlated errors. 

This work demonstrates all ingredients and the operation of the self-referenced current source, at a repetition frequency of about 30 Hz, i.e. in the limit $f\ll 1/\tau_d$. The resulting current of our device is limited here to about $4.8$~aA. When operated at higher repetition frequencies above the detector bandwidth, the ratio of the average time between errors $((1-P_1)\cdot f)^{-1}$ to the detector bandwidth $1/\tau_d$ has to be included in the statistical analysis. Then the remaining uncertainty after correction $u_c$ owing to potential misattributions can be approximated by~\cite{Wulf2013} $$u_c\approx \frac{2N!}{\left(\frac{N+1}2\right)!\left(\frac{N-1}2\right)!}\left(1-P_1\right)^{\frac{N+1}2}\left(f\cdot \tau_d\right)^{\frac {N-1}2}$$ with $N$ the number of pumps in series and $P_1$ the average probability of all pumps to transfer exactly one electron per cycle. In the case of $N=3$ as used here this formula simplifies to $u_c\approx 6(1-P_1)^2\cdot f\cdot \tau_d.$

To evaluate the present technical limits of this technique we assume a circuit with five serial pumps and an increased detector bandwidth of $1/\tau_d\approx 50$~kHz which can be obtained using RF-SET technology~\cite{Schoelkopf1998}. Assuming further a pumping frequency of $f=1$~GHz and a pump error probability of $|1-P_1| \approx 1\cdot 10^{-6}$ as experimentally demonstrated~\cite{Giblin2012} a very low relative uncertainty of the corrected output current of $u_c<10^{-8}$ could be achieved. Via the quantum metrological triangle~\cite{Likharev1985,Scherer2012}, such a device is required for a direct comparison of a well characterized single-electron output current with the Quantum Hall and the Josephson effect to lay the foundation for the future redefinition of the electrical SI units in terms of natural constants. 

We can conclude that the concept of error accounting~\cite{Wulf2013} as implemented in our self-referenced single-electron current source is a feasible route for metrologically accurate, quantized current generation using imperfect current sources suffering from the intrinsic stochastic nature of quantum-mechanical tunnelling. Combined with highly precise current scaling and measurement techniques~\cite{Sullivan1974,Giblin2012} such a single-electron current standard would close the conceptual gap between macroscopic- and \emph{single-}electronics~\cite{Likharev1999}.

We thank N. Ubbelohde and A. Zorin for valuable discussion and support and acknowledge support in clean-room processing by P. Hinze. This work has been supported by the DFG and within the Joint Research Project "Quantum Ampere" (JRP SIB07) within the European Metrology Research Programme (EMRP). The EMRP is jointly funded by the EMRP participating countries within EURAMET and the European Union.


\clearpage

\begin{figure}[htp] \centering
\centering \includegraphics[]{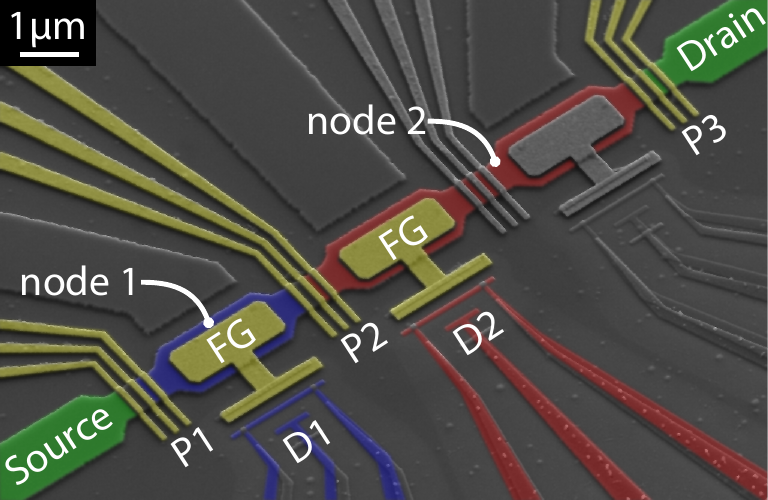} \caption{False-color SEM image of the device under investigation. The semiconductor part of the device between source and drain (green) consists of three pumps P1 - P3 and two charge nodes 1,2 (blue, red) in between. Each pump is defined by three metallic top gates (yellow) defining a QD in the semiconductor channel. Between P2 and P3, another pump structure (grey) is visible, which is not used in the experiment and electrically grounded. In the lower part, the single-electron transistor (SET) detectors are shown (D1 and D2, coloured blue and red, respectively). These are capacitively coupled via metallic floating gates (FG, yellow) to the two charge nodes allowing to detect the node charge on the single-electron level.
}\end{figure}

\begin{figure}[htp] \centering
\centering \includegraphics{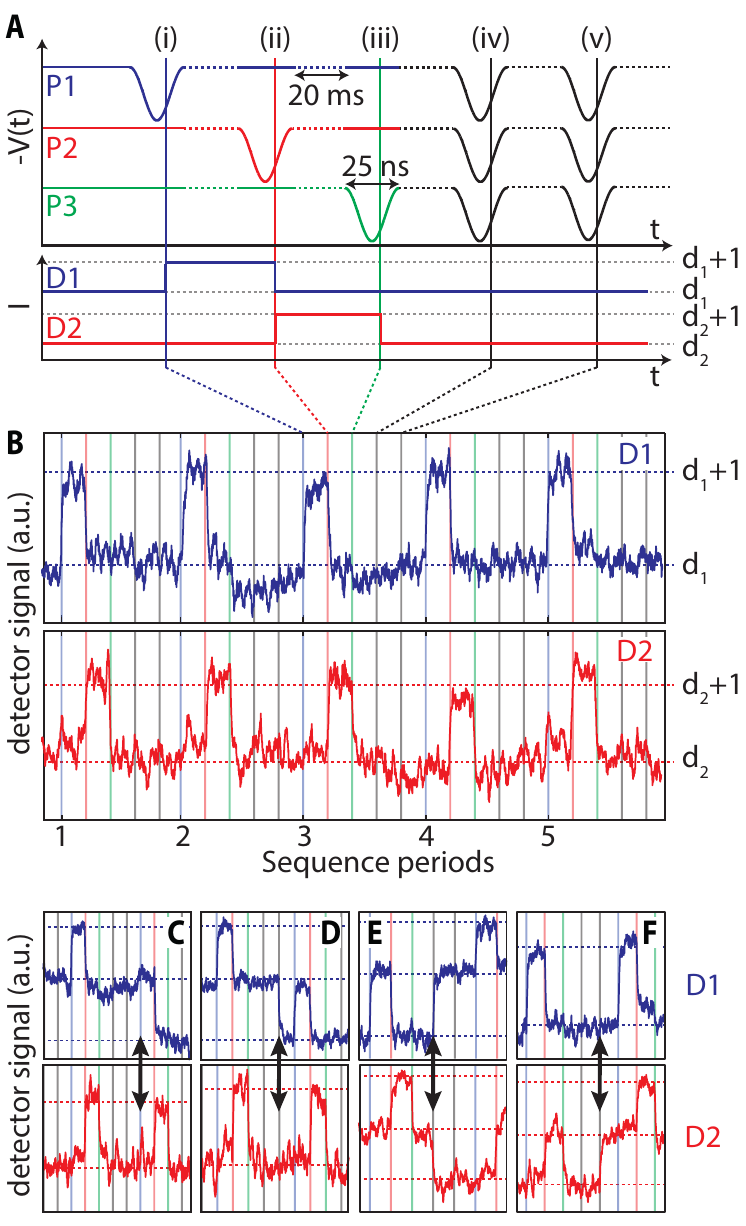} \caption{Operation of the self-referenced current source. {\bf A} Sketch of the pulse sequence for pumps P1, P2 and P3 as well as the corresponding nominal detector signals for error free pumping of one electron per pulse. The pumps are set to transfer one electron per pulse. A marker sequence (pulses (i)-(iii)) shuttles one electron through the structure sequentially, allowing for an on-the-fly self-calibration of the detectors in terms of electrons on the nodes. The outcome of such a sequence is reflected by the step-like detector response shown in the bottom of this sketch. Subsequently, all three pumps are triggered twice simultaneously (pulses (iv)-(v)). Instead of observing transferred electrons as in the marker sequence, we only monitor transfer errors here. All pulses are delayed by $20\ $ms each. {\bf B} Corresponding measured signals for the pulse sequence shown in A. Vertical lines indicate the different pump pulses, horizontal dashed lines show the two charge states $d_{(1,2)},d_{(1,2)}+1$  of each node $1,2$ as a guide to the eye. {\bf C-F} Signatures of missing-cycle events marked by double arrows: {\bf C} During the marker sequence by pump P1, {\bf D-F} in series operation by pump P1, P2 and P3, respectively.}
\end{figure}

\begin{figure*}[htp] \centering
\includegraphics[]{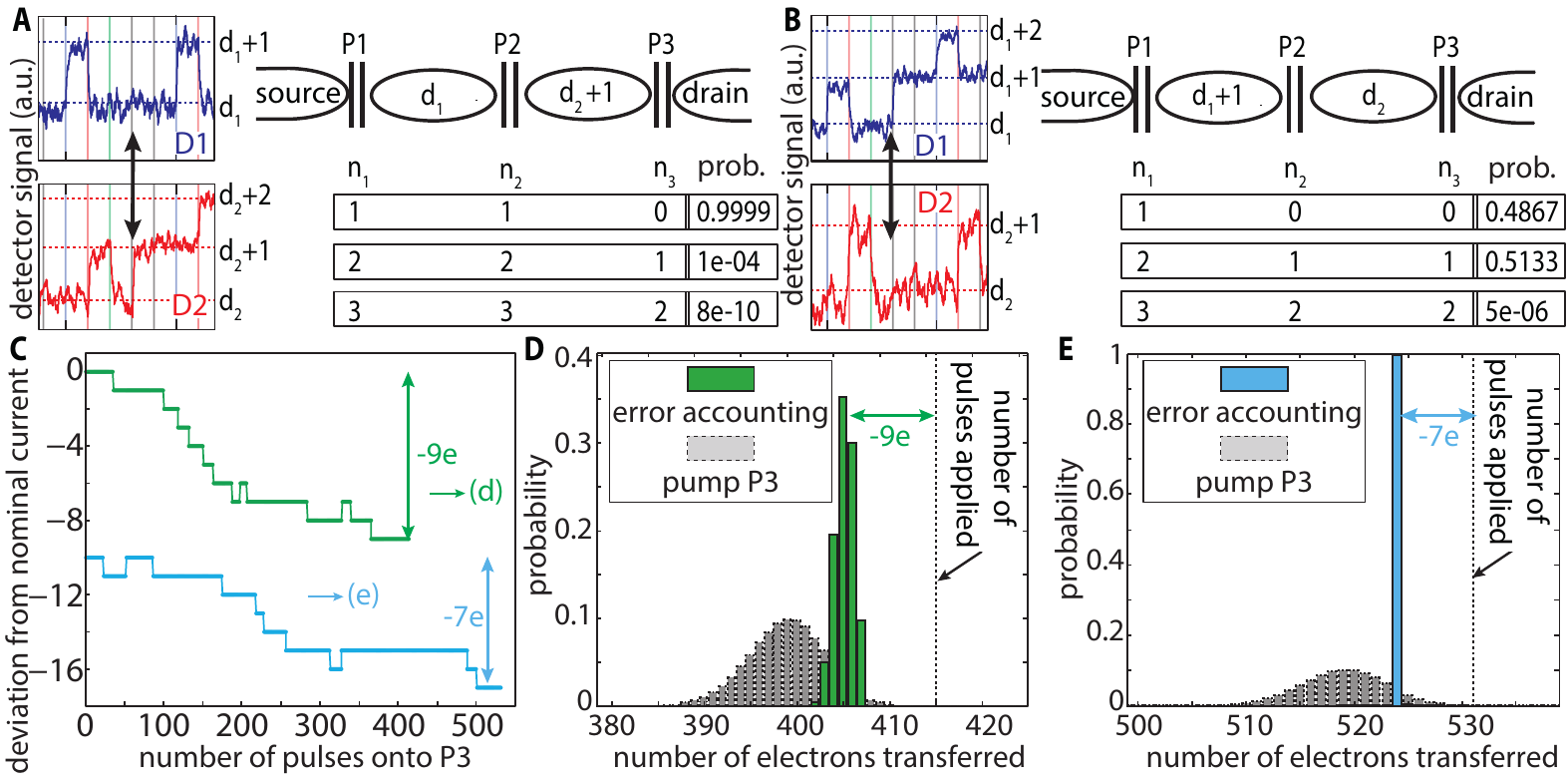} \caption{Error analysis. {\bf A} Error signature $(d_1,d_2+1)$ measured (left) and sketched (right) with different scenarios of realization. Additionally, the probability vector is shown. {\bf B} Error signature $(d_1+1,d_2)$ measured (left) and sketched (right) with most likely numbers of electrons transferred by the pumps. {\bf C} Deviation from nominal current across P3 as a function of pump pulses for two different working points of the pumps. The length of the traces is limited by the requirement on both detectors D1 and D2 to be sensitive. Curves are offset by $-10$ for clarity. {\bf D} Probability distribution of the number of pumped electrons of the green trace in C. The result of the error-accounting scheme (green, solid) is compared to the expected distribution for an individual pump P3 (grey, dashed lines). The dashed vertical line corresponds to the number of pump pulses applied to P3 (equal to the nominal number of electrons transferred). {\bf E} Comparison of probability distributions as in D for the blue plot in C between the individual pump P3 and the self-referenced current source.}
\end{figure*}

\clearpage

\section*{Supplement}
\setcounter{figure}{0}
\renewcommand{\thefigure}{S.\arabic{figure}}

\subsection*{Measurement setup}
All measurements are performed in a dry dilution refrigerator at nominal base temperature of about $25$~mK. The dc-lines are low-pass filtered and thermally coupled at the mixing chamber using self-made copper-powder filters~\cite{Martinis1987}. All coaxial lines providing rf-signals to drive the pumps are low-pass filtered at $80$~MHz limiting the maximum driving frequency of the pump pulses. We use synchronized signal generators (Agilent M9330A/9331A) and digitizers (NI-PXI 5105) installed in a common NI PXI frame.  All detector signals are amplified by home-made transimpedance amplifiers (Gain of $100$~M$\Omega$) and digitized at $6$~MS/s, digitally averaged to $12$~kS/s and subsequently saved to hard disk. The time resolution of our electron counting experiment is limited by the RC rise time of the amplifier response at this gain.

\subsection*{Detector calibration}
The SETs used here are operated at fixed bias voltage, the current varying as a function of external potentials is measured. This periodic SET response for a fixed voltage bias is shown in Figs. S.1A and S.1C for detector D1 and D2, respectively. Instead of creating an electrostatic potential on the semiconducting nodes by adding electrons, we continuously change the voltage on the SETs' gates here. To cope with the periodicity, we divide the SET response in two parts for positive and negative slope, respectively, and analyze the response during the marker sequence separately for both parts. At the extrema (above and below the dashed horizontal lines), the SETs are insensitive to variations in the external potentials, finally limiting the observation time. Due to the unidirectionality of the pump's charge transfer and the exact timing between digitizers and pump-pulse generators, we are able to identify the jumps observed in the detector traces as surplus or missing electrons. This knowledge also enables us to determine whether we measure at the rising or the falling edge of the corresponding detector.

As an example, during the marker sequence we expect to add electrons to node 1 by the first pulse (via P1) and to remove electrons by the following pulse to node 2 via pump P2. If the detector current shifts upwards after the first pulse (electrons added to node 1) and back down after the second pulse, we know for certain that the SET is operated on the falling edge. Since the pumps' working points are chosen close to $n_i=1$, we expect to transfer mainly one electron per cycle. By creating pairs of SET current before the pulse and the corresponding response after the pulse, we obtain a frequency distribution of the response as a function of SET working point which should show maxima of the response at $\pm 1$ electron.

\begin{figure}[htp] \centering
\centering \includegraphics[]{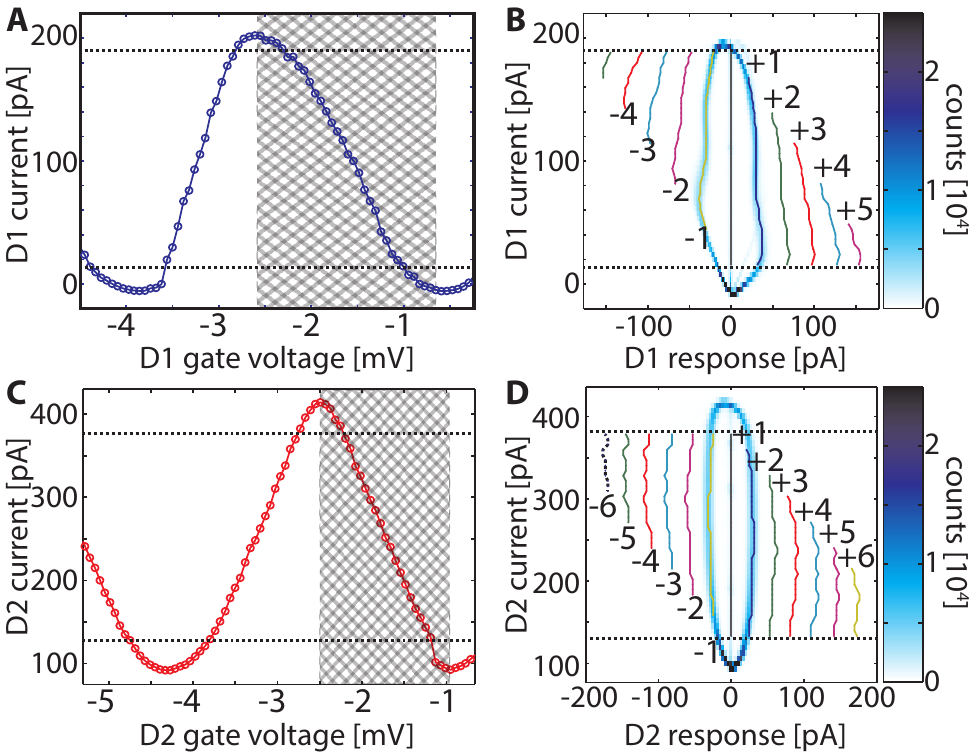} \caption{
Calibration of the SETs. {\bf A} Current across D1 for a fixed bias voltage as a function of the SET's gate. The periodicity of the response is about $3$~mV. The calibration is divided into two parts for rising and falling slopes. The latter is emphasized by the shaded area. {\bf B} Pseudocolour plot of the SET response on the falling edge (shaded area in A) of D1 during the marker sequence. The y-axis corresponds to the one in A. The colour code shows the frequency of an SET response to a marker pulse for a given working point. Negative response corresponds to removing electrons from the node, positive one to adding electrons. The dashed horizontal lines indicate the limits of sensitivity. Below and above, the detector is considered to be insensitive. The dots and lines show the predictions to add/remove up to six electrons. {\bf C} Same as A for D2. The periodicity of the response is also about $3$~mV. {\bf D} Analysis of the response of D2 during the marker sequence as in B.
}\end{figure}

The resulting frequency distribution is shown in Figs. S.1B (D1) and S.1D (D2) as a pseudocolour plot for the falling edge of both detectors. The y-axis corresponds to the axis in A (D1) and C (D2), respectively. The x-axis shows the SET response, i.e. the change in SET current for a distinct change in the number of electrons on the node. The colour reflects the frequency of the response for the given working point according to the colour bar. The dashed horizontal lines again reflect the limits of sensitivity. Zero response at the center of the x-axis divides the plot two parts: On the left hand side (negative response), we show the response for removing, on the right hand side (positive response) the response for adding electrons to the node. Clear maxima for $\pm 1$ electron on the node are visible. By detecting these maxima and using these as a reference, we are able to identify changes of electrons on the node of up to $\pm 6$. The predicted changes based on this analysis are shown by the dots and lines. The lower the SET working point, the more (less) electrons added (removed) can be detected due to the limitations of the dynamic range.

By comparing the SET response for an arbitrary pulse at a given working point to these predictions, we directly convert the signature into a discrete number of electrons. Fig. S.2 shows the real-time data (A) as well as the counting signal (B) for a longer trace of 25 periods (i.e., for a length of $2.5$~s).

\begin{figure}[htp] \centering
\centering \includegraphics[]{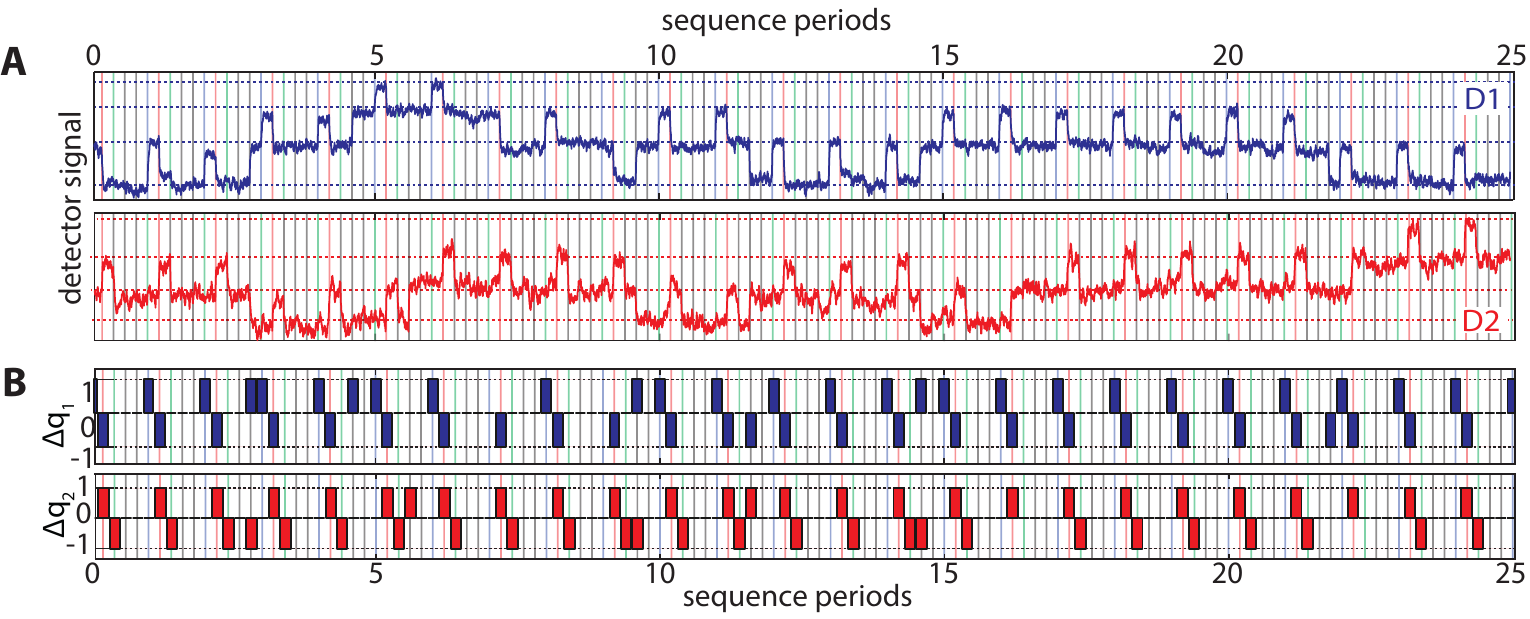} \caption{
Real-time data and counting signal. {\bf A} Real-time data for both detectors D1 (upper graph, blue) and D2 (lower graph, red) for 25 sequence periods. {\bf B} Corresponding counting signal for both detectors.
}\end{figure}

\subsection*{Full counting statistics of the pumps during the marker sequence}
For a given working point of the pumps, set by applying a negative voltage to the exit gate of each pump, we analyze the pumps' working point by counting the number of electrons transferred during the marker sequence. Here, we remove all sequences in which the corresponding detector observing the node attached to the pump was insensitive. For pump P2, connecting both nodes, we analyze both detectors independently. The result contains the counting statistics of each pump. Due to the limited number of observations (we measure at each working point for 10 minutes, which corresponds to about 6000 sequence periods, but the detectors are not always sensitive as explained above), we estimate an upper bound of the unobserved transfer rates by the inverse number of total marker events for the specific pump and working point.

\end{document}